\documentclass[aps,prd,superscriptaddress,showkeys,showpacs,twocolumn,a4paper]{revtex4-1}
\usepackage{ulem}
\usepackage{color}
\usepackage{graphicx}
\usepackage{appendix}
\usepackage{epsfig}
\usepackage[colorlinks=true,linkcolor=blue]{hyperref}
\usepackage{epstopdf}
\usepackage{amsmath}
\usepackage{subfig}
\usepackage{comment}
\usepackage[dvipsnames]{xcolor}
\newcommand{\be}{\begin{equation}}
\newcommand{\ee}{\end{equation}}
\newcommand{\ba}{\begin{eqnarray}}
\newcommand{\ea}{\end{eqnarray}}

\begin{document}
 
    \title{Dragging heavy quark in an anisotropic QCD medium beyond the static limit} 
\author{Avdhesh Kumar}
\email{avdhesh.5000@gmail.com}
\affiliation{Indian Institute of Technology Gandhinagar, Gandhinagar-382355, Gujarat, India}

\author{Manu Kurian}
\email{manu.kurian@iitgn.ac.in}
\affiliation{Indian Institute of Technology Gandhinagar, Gandhinagar-382355, Gujarat, India}

\author{Santosh K. Das}
\email{santosh@iitgoa.ac.in}
\affiliation{School of Physical Sciences, Indian Institute of Technology Goa, Ponda-403401, Goa, India}

\author{Vinod Chandra}
\email{vchandra@iitgn.ac.in}
\affiliation{Indian Institute of Technology Gandhinagar, Gandhinagar-382355, Gujarat, India}
	\date{\today} 
	\bigskip
	\begin{abstract}
Heavy quark dynamics in an anisotropic QCD medium have been analyzed within the Fokker-Planck approach. Heavy quark drag force and momentum diffusion tensor have been decomposed by employing a general tensor basis for an anisotropic medium.
Depending upon the relative orientation of the direction of the momentum anisotropy of the medium and heavy quark motion, two drag and four diffusion coefficients have been estimated in the anisotropic QCD medium. The relative significance of different components of drag and momentum diffusion coefficients has been explored. The dependence of the angle between the anisotropic vector and heavy quark motion to the drag and diffusion coefficients has also been studied. Further, the energy loss of heavy quarks due to the elastic collisional process in an anisotropic medium has been studied. It is seen that the anisotropic contributions to heavy quark transport coefficients and its collisional energy loss have a strong dependence on the direction and strength of momentum anisotropy in the QCD medium. 
	 
	\end{abstract}
     
\date{\today}

	
	
\maketitle
%
%
\section{Introduction}
 
The heavy-ion collision experiments pursued at the Relativistic Heavy Ion Collider (RHIC) and the Large Hadron Collider (LHC) have confirmed the existence of strongly interacting matter, the quark-gluon plasma (QGP)~\cite{Adams:2005dq, Back:2004je,Arsene:2004fa,Adcox:2004mh,Aamodt:2010pb,Jaiswal:2020hvk}.  Among various signatures from experimental observable, heavy quarks (HQs), mainly charm and bottom quarks, are identified as the excellent experimental probe to study the properties of the hot QCD medium~\cite{Rapp:2018qla,Dong:2019unq,Prino:2016cni,Aarts:2016hap,Andronic:2015wma,Das:2010tj,Das:2013kea,Moore:2004tg,Gossiaux:2006yu,Brambilla:2020siz,Rapp:2009my}. HQs undergo random motion and witness the QCD medium expansion. This is attributed to the fact that HQs are mostly created in the early stages of heavy-ion collision, and their thermalization time is greater than the lifetime of the QGP. Several theoretical efforts have been made to explore HQ dynamics and the associated experimental observables such as nuclear suppression factor $R_{AA}$, flow coefficients~\cite{vanHees:2005wb,Gossiaux:2008jv,Das:2009vy,He:2012df,Cao:2013ita,Cao:2018ews,Xu:2018gux,Das:2015ana,Scardina:2017ipo,Cao:2016gvr,Song:2015sfa,Adare:2006nq,Adler:2005xv,Alberico:2013bza,vanHees:2007me,Li:2019wri,Akamatsu:2008ge,Uphoff:2011ad,Xu:2013uza,Zigic:2018ovr}. A few attempts have been done to explore the impact of momentum anisotropic aspects of the QCD medium on HQ transport. However, a systematic study of HQ transport by constructing the drag force and diffusion tensor using a general tensor basis in an anisotropic medium is essential for the proper understanding of HQ observables.

Momentum anisotropy arises due to the rapid expansion of the created QCD medium in the longitudinal direction compared to the transverse directions and may sustain in the entire evolution of the medium. This anisotropy may induce instability to the Yang-Mills fields (Chromo-Weibel instability) and may have a vital role in the evolution of the QCD medium~\cite{Mrowczynski:1993qm,Randrup:2003cw,Chandra:2012qq,Chandra:2011bu}. It has been argued that the QCD medium has an anomalous viscosity that arises from Chromo-Weibel instabilities, which may provide a possible explanation for the near-perfect liquidity of the QGP without considering the strongly coupled state assumption~\cite{Asakawa:2006tc,Asakawa:2006jn}. The momentum anisotropic aspects have been explored in the context of electromagnetic probes~\cite{Shen:2014nfa,Kasmaei:2019ofu,Kasmaei:2018oag}, collective modes of QCD~\cite{Romatschke:2003ms,Schenke:2006xu, Kumar:2017bja} and in the hydrodynamical expansion of the medium~\cite{Alqahtani:2017mhy,Alqahtani:2017tnq,Alqahtani:2017jwl}. Medium anisotropy will affect the dynamics of HQ and can be quantified in terms of its transport coefficients, drag and momentum diffusion, in the medium. A magnetic field-induced anisotropy to the HQ momentum diffusion has been recently explored in Refs.~\cite{Fukushima:2015wck,Singh:2020faa} and generates huge attention towards the recent RHIC and LHC observations~\cite{ALICE:2019sgg,STAR:2019clv}. Non-equilibrium effects of the QCD medium to the HQ transport have been studied in Refs.~\cite{Kurian:2020orp,GolamMustafa:1997id,Srivastava:2016igg, Song:2019cqz,Prakash:2021lwt,Shaikh:2021lka, Romatschke:2004au,Das:2012ck}. 

The focus of the present study is to set up a general framework to explore HQ dynamics in an anisotropic QCD medium for arbitrary relative orientation of the direction of anisotropy and HQ motion. To that end, we have performed a decomposition of HQ drag force and momentum diffusion tensor using a tensor basis for an anisotropic medium. This gives rise to two drag and four diffusion coefficients of HQ in the QCD medium. The anisotropic effects are entering through the non-equilibrium part of the distribution and are obtained by solving the transport equation. We have analyzed the impact of anisotropy on the temperature and momentum dependence of the HQ transport coefficients in the medium. In addition, we have explored the dependence of the orientation of HQ motion with the direction of anisotropy on HQ dynamics in the medium.  These anisotropic transport coefficients may have a significant role in the estimation of HQ experimental observables in the heavy-ion collision experiments by treating it as input parameters in the Langevin dynamics.  

The manuscript is organized as follows. Section~\ref{sec:2} is devoted to the theoretical formulation of HQ transport  along with the general decomposition of drag and momentum diffusion tensor in an anisotropic QCD medium. In section~\ref{sec:3}, we have presented the results of HQ transport coefficients and its collisional energy loss in the anisotropic medium. We have summarized the analysis with an outlook in section~\ref{sec:4}.

{\it{Notations and conventions}}: In the manuscript, the subscript $k$ represents the particle species of the medium, $i.e.$, $k=(g, \Tilde{q})$ with $g$ and $\Tilde{q}$ denote the gluons and quarks. HQ energy is defined by  $E_p=\sqrt{\mid{\bf{p}}\mid^2+m_{HQ}^2}$ where ${\bf{p}}$ and $m_{HQ}$ respectively denote the momentum and mass of HQ. Energy of constituent particles (in the massless limit) is represented as $E_q=\mid{\bf{q}}\mid$ with ${\bf{q}}$ as the momentum. 
The quantity $a_k=1, -1, 0$ for Bose-Einstein, Fermi-Dirac and Maxwell-Boltzmann distributions, respectively. 
\section{ Heavy quark Drag and diffusion }\label{sec:2}
The dynamics of HQ in the hot QCD medium is considered as Brownian motion and can be described within the Fokker-Planck equation as follows~\cite{Svetitsky:1987gq,GolamMustafa:1997id},
\begin{align}\label{1.1}
  	\frac{\partial f_{HQ}}{\partial t}=\frac{\partial}{\partial p_i}\left[A_i({\bf p})f_{HQ}+\frac{\partial}{\partial p_j}\Big(B_{i j}({\bf p})f_{HQ}\Big)\right],
  	\end{align}
where $f_{HQ}$ denotes the HQ distribution in the medium. The interactions of HQ with the light quarks and gluons are quantified in terms of drag force $A_i$ and momentum diffusion $B_{ij}$ in the QGP medium. In the current analysis, we consider the two-body elastic collisional process $HQ(P)+l(Q)\rightarrow HQ(P^{'})+l(Q^{'})$, where $l$ denotes quarks/antiquarks, and gluons. Here, $P=(E_p,\bf{p})$ and $Q=(E_q,\bf{q})$ define the four-momentum of HQ and medium constituent particle before the interaction. The matrix element $|\mathcal{M}_{2\rightarrow 2}|$ for the  elastic collisions of HQs with medium particles has been investigated in Ref.~\cite{Combridge:1978kx, Svetitsky:1987gq}. The drag force of HQ describes the thermal average of the momentum transfer due to the interaction, whereas the momentum diffusion quantifies the average of the square of the momentum transfer. The HQ drag and momentum diffusion in the QGP medium take the forms as follows,
 \begin{align}\label{1.2}
    A_i=&\frac{1}{2E_p}\int{\frac{d^3{\bf q}}{(2\pi)^32E_q}}\int{\frac{d^3{\bf q}'}{(2\pi)^32E_{q'}}}\int{\frac{d^3{\bf p}'}{(2\pi)^32E_{p'}}}\frac{1}{\gamma_{HQ}}\nonumber\\ &\times \sum|\mathcal{M}_{2\rightarrow 2}|^2(2\pi)^4\delta^4 (P+Q-P'-Q') f_{k}({\bf{q}}) \nonumber\\ &\times \Big(1+a_k f_{k}({\bf{q'}})\Big)\Big[({\bf p}-{\bf p}')_i\Big]=\langle\langle({\bf p}-{\bf p}')_i\rangle\rangle,
\end{align}
\begin{align}\label{1.3}
    B_{ij}=\frac{1}{2}\langle\langle({\bf p}-{\bf p}')_i({\bf p}-{\bf p}')_j\rangle\rangle,
\end{align} 
where $\gamma_{HQ}$ is the statistical degeneracy factor of the HQ, $f_{k}$ represents the near-equilibrium distribution function of quark/antiquark and gluon.   
In general, HQ drag and diffusion coefficients can be schematically described as,
\begin{align*}
    X_c=\int \text{phase space}\times \text{ interaction}\times \text{transport part}.
\end{align*}
HQ transport coefficients can be obtained with the proper decomposition of drag force and momentum diffusion matrix in the background QGP medium. We proceed with the decomposition of $A_i$ and $B_{ij}$ in the isotropic QCD medium.
%
\subsection{\label{subsec:2.1} For isotropic QCD medium } 
%
In an isotropic medium, the drag force depends on the HQ momentum and $A_{i}$ can be decomposed as,
\begin{align}\label{1.5}
A_{i}=p_{i}A_0(p^2), 
\end{align}
where $p^2=|{\bf p}|^2$ and $A_0$ is the drag coefficient of the HQ in the isotropic QGP medium.
The drag coefficient can be obtained from Eq.~(\ref{1.2}) and Eq.~(\ref{1.5}) as,
\begin{align}\label{1.6}
A_0=p_{i}A_{i}/p^2=\langle\langle 1 \rangle\rangle - \frac{\langle\langle {\bf{p\cdot p'} \rangle\rangle}}{p^2}. 
\end{align}
Similarly, $B_{ij}$ can be decomposed into longitudinal and transverse components in the isotropic QCD medium as, 
\begin{align}\label{1.7}
&B_{ij} = \left(\delta_{ij}-\frac{p_ip_j}{p^2}\right) B_0(p^2)+\frac{p_ip_j}{p^2}B_1(p^2),
\end{align}
where the transverse and longitudinal diffusion coefficients can be defined as follows,
\begin{align}\label{1.8}
&B_0=\frac{1}{4}\left[\langle\langle p'^{2} \rangle\rangle-\frac{\langle\langle ({\bf{p'\cdot p}})^2\rangle\rangle}{p^2} \right],\\ 
&B_1= \frac{1}{2}\left[\frac{\langle\langle ({\bf{p'\cdot p})}^2\rangle\rangle}{p^2} -2\langle\langle ({\bf{p'\cdot p})}\rangle\rangle +p^2 \langle\langle 1 \rangle\rangle\right]\label{1.9}.
\end{align}
The kinematics of $2\rightarrow 2$ process can be simplified in the center-of-momentum (COM) frame of the system, and the average of a function $F({\bf p})$ in the COM frame for the isotropic medium can be described as follows,
\begin{align}\label{1.10}
    \langle \langle F({\bf p})\rangle \rangle=&\frac{1}{(512 \, \pi^4) E_p \gamma_{HQ}}\int_0^\infty dq \left(\frac{s-m_{HQ}^2}{s}\right) f^{0}_k(E_q)   \nonumber\\ &\times \int_0^\pi d\chi \, \sin\chi\int_0^\pi d\theta_{cm} \, \sin\theta_{cm}\sum{|{\mathcal{M}}_{2\rightarrow2}|^2} \nonumber\\ &\times \int_0^{2\pi} d\phi_{cm} (1+a_k f^{0}_k(E_{q'})) \ F({\bf p}),
\end{align}
where $f_k^{0}$ is the isotropic distribution function, and $\chi$ quantifies the angle between the incident  HQ and medium constituent particles in the lab frame. The quantities $\theta_{cm}$ and $\phi_{cm}$ respectively describe the zenith and azimuthal angle in the COM frame. Here, the Mandelstam variables $s,t,u$ are defined as follows,
\begin{align}\label{1.11}
s = & (E_p+E_q)^2-(p^2+q^2+2 p q \cos \chi), \\
t = & 2 \, p_{cm}^2 (\cos \theta_{cm}-1),\\
u = & 2 \, m_{HQ}^2-s-t,
\end{align}
with $p_{cm}=|{\bf p}_{cm}|$ as the magnitude of initial momentum of HQ in the COM frame. 
\subsection{\label{subsec:2.2} For an anisotropic QCD medium }
Momentum anisotropies arise due to the rapid expansion of the hot QCD medium in the early stages of the relativistic heavy-ion collisions. In the present analysis, the impact of momentum anisotropy is entering through the distribution function of the medium constituent particles. The anisotropic momentum distribution can be described in terms of isotropic distribution function by re-scaling one direction in momentum space as follows~\cite{Romatschke:2003ms,Schenke:2006xu},
\begin{align}\label{1.11}
    f_k^{{\text{(aniso)}}}({\bf q})= \sqrt{1+\xi}\,f_k^{0}\Big(\sqrt{q^2+\xi({\bf q}\cdot{\bf n})^2}\Big),
\end{align}
where $\xi$ is the anisotropic parameter that quantifies the stretching or squeezing of the momentum distribution in the prescribed direction ${\bf n}$ where ${\bf n}$ is the unit vector that indicates the direction of momentum anisotropy in the medium. The present focus is on a weakly anisotropic medium such that $\xi\ll 1$ and the distribution function reduces to the form $f_k^{{\text{(aniso)}}}({\bf q})= f_k^{{0}}+\delta f_k$ with~\cite{Srivastava:2015via},

\begin{align}\label{1.12}
   \delta f_k=-\frac{\xi}{2 E_q T}({\bf q}\cdot{\bf n})^2 (f^{0}_k)^2\exp{\Big(\frac{E_q}{T}\Big)}.
\end{align}
%
By defining $\Tilde{n}^i=\left(\delta_{ij}-\frac{p_ip_j}{p^2}\right)n^j$ such that ${\bf p}\cdot{\bf \Tilde{n}}=0$, the drag force in the anisotropic medium can be decomposed on the orthogonal basis as follows,
\begin{align}\label{1.13}
A_{i}=p_{i}A^{\text{(aniso)}}_0+\Tilde{n}_{i}A^{\text{(aniso)}}_1.
\end{align}
The components of the drag force in the anisotropic medium can be obtained as follows,
\begin{align}\label{1.14}
&A^{\text{(aniso)}}_0=p_{i}A_{i}/p^2=\langle\langle 1 \rangle\rangle - \frac{\langle\langle {\bf{p\cdot p'} \rangle\rangle}}{p^2},\\
& A^{\text{(aniso)}}_1=\Tilde{n}_{i}A_{i}/\Tilde{n}^2= - \frac{1}{\Tilde{n}^2}\langle\langle {\bf{\Tilde{n}\cdot p'} \rangle\rangle},
\end{align}
where {$ \Tilde{n}^2=1-\frac{({\mathbf{p}}\cdot{\mathbf{\hat{n}}})^2}{p^2}=1-\cos^2\theta_{n}$}. 
Employing the definition of the near-equilibrium distribution function as described in Eq.~(\ref{1.12}), the average of a function $F(p')$ in the anisotropic medium can be defined as,
\begin{align}\label{1.15}
   \langle\langle F({p'})\rangle\rangle=\langle\langle F({p'})\rangle\rangle_0+\langle\langle F({p'})\rangle\rangle_{\text{a}},
\end{align}
where the isotropic part $\langle\langle F({p'})\rangle\rangle$ is defined in Eq.~(\ref{1.10}). 
Following the same prescription as in the case of isotropic case, we can represent $\langle\langle F({p'})\rangle\rangle_{a}$ in the COM frame as,
\begin{align}\label{1.151}
    &\langle \langle F({\bf p})\rangle \rangle_a=\frac{1}{(1024 \, \pi^5) E_p \gamma_{HQ}}\int_0^\infty dq q\left(\frac{s-m_{HQ}^2}{s}\right) \nonumber\\ &\times\int_0^\pi d\chi \, \sin\chi \int_0^{2\pi} d\phi  \int_0^\pi d\theta_{cm} \, \sin\theta_{cm}\sum{|{\mathcal{M}}_{2\rightarrow2}|^2}\nonumber\\ &\times \int_0^{2\pi} d\phi_{cm} \ \Big[\delta f_k({\bf{q}})\Big(1+a_k f^0_{k}({\bf{q'}})\Big) +a_k f^0_{k}({\bf{q}})\delta f_{k}({\bf{q'}})\Big] F({\bf p}).
\end{align}
Employing Eq.~(\ref{1.15}) in  Eq.~(\ref{1.14}), we obtain the non-equilibrium correction to the HQ drag coefficient described in Eq.~(\ref{1.6}) as,
\begin{align}
A^{\text{(aniso)}}_0= A_0+ \delta A_{0},
\end{align}
where $A_0$ is the isotropic part and $\delta A_{0}$ represents the anisotropic corrections to the drag coefficient in the QGP medium and can be obtained from Eq.~(\ref{1.151}). Further, the term $A^{\text{(aniso)}}_1$ is the additional component of the drag coefficient that arises due to the anisotropy of the medium.  
To decompose the HQ diffusion, one needs to construct the appropriate tensor basis for the symmetric matrix $B_{ij}$ with the momentum vector $p^i$ and anisotropy vector $n^i$.  Following Ref.~\cite{Romatschke:2003ms}, we decompose the $B_{ij}$ into four components as follows,
\begin{align}\label{1.16}
B_{ij} =& \left(\delta_{ij}-\frac{p_ip_j}{p^2}\right) B^{\text{(aniso)}}_0+\frac{p_ip_j}{p^2}B^{\text{(aniso)}}_1\nonumber\\&+\frac{\Tilde{n}_i\Tilde{n}_j}{\Tilde{n}^2}B^{\text{(aniso)}}_2+(p^i\Tilde{n}^j+p^j\Tilde{n}^i)B^{\text{(aniso)}}_3,
\end{align}
The components of the momentum diffusion can be obtained by taking the appropriate projections of the Eq.~(\ref{1.16}) and have the following forms,
\begin{widetext}
\begin{align}
   &B^{\text{(aniso)}}_0= \bigg[\left(\delta_{ij}-\frac{p_ip_j}{p^2}\right)-\frac{\Tilde{n}_i\Tilde{n}_j}{\Tilde{n}^2}\bigg]B_{ij}=\frac{1}{2}\left[\langle\langle p'^{2} \rangle\rangle-\frac{\langle\langle ({\bf{p'\cdot p}})^2\rangle\rangle}{p^2}-\frac{\langle\langle ({\bf{p'\cdot \Tilde{n}}})^2\rangle\rangle}{\Tilde{n}^2} \right],\label{B0.0}
   \end{align}
   \begin{align}
   &B^{\text{(aniso)}}_1=\frac{p_ip_j}{p^2}B_{ij}= \frac{1}{2}\left[\frac{\langle\langle ({\bf{p'\cdot p})}^2\rangle\rangle}{p^2} -2\langle\langle ({\bf{p'\cdot p})}\rangle\rangle +p^2 \langle\langle 1 \rangle\rangle\right],\label{B0.1}\\
   &B^{\text{(aniso)}}_2= \bigg[\frac{2\Tilde{n}_i\Tilde{n}_j}{\Tilde{n}^2}-\left(\delta_{ij}-\frac{p_ip_j}{p^2}\right)\bigg]B_{ij}=\frac{1}{2}\left[-\langle\langle p'^{2} \rangle\rangle+\frac{\langle\langle ({\bf{p'\cdot p}})^2\rangle\rangle}{p^2}+\frac{2\langle\langle ({\bf{p'\cdot \Tilde{n}}})^2\rangle\rangle}{\Tilde{n}^2} \right],\\
   &B^{\text{(aniso)}}_3=\dfrac{1}{2p^2\Tilde{n}^2} (p^i\Tilde{n}^j+p^j\Tilde{n}^i) B_{ij}=\dfrac{1}{2p^2\Tilde{n}^2}\left[-p^2\langle\langle({\bf{p'\cdot \Tilde{n}}})\rangle\rangle+\langle\langle ({\bf{p'\cdot p}})({\bf{p'\cdot \Tilde{n}}})\rangle\rangle \right]\label{B0.00}.
\end{align}
\end{widetext}
Note that Eqs.~(\ref{B0.0})-(\ref{B0.00}) will reduce back to the results of Ref.~\cite{Srivastava:2016igg} in the case of $\langle\langle {\bf{\Tilde{n}\cdot p'}}\rangle\rangle=0$. Note that we have obtained $\langle\langle {\bf{\Tilde{n}\cdot p'}}\rangle\rangle=0$ for the isotropic case. However, for the anisotropic medium, by employing the general tensor decomposition, we observe that the term $\langle\langle {\bf{\Tilde{n}\cdot p'}}\rangle\rangle_a$ is non-zero and modify the HQ transport coefficients.

\begin{figure}
    \centering
    \hspace{-.5cm}
    \includegraphics[width=0.48\textwidth]{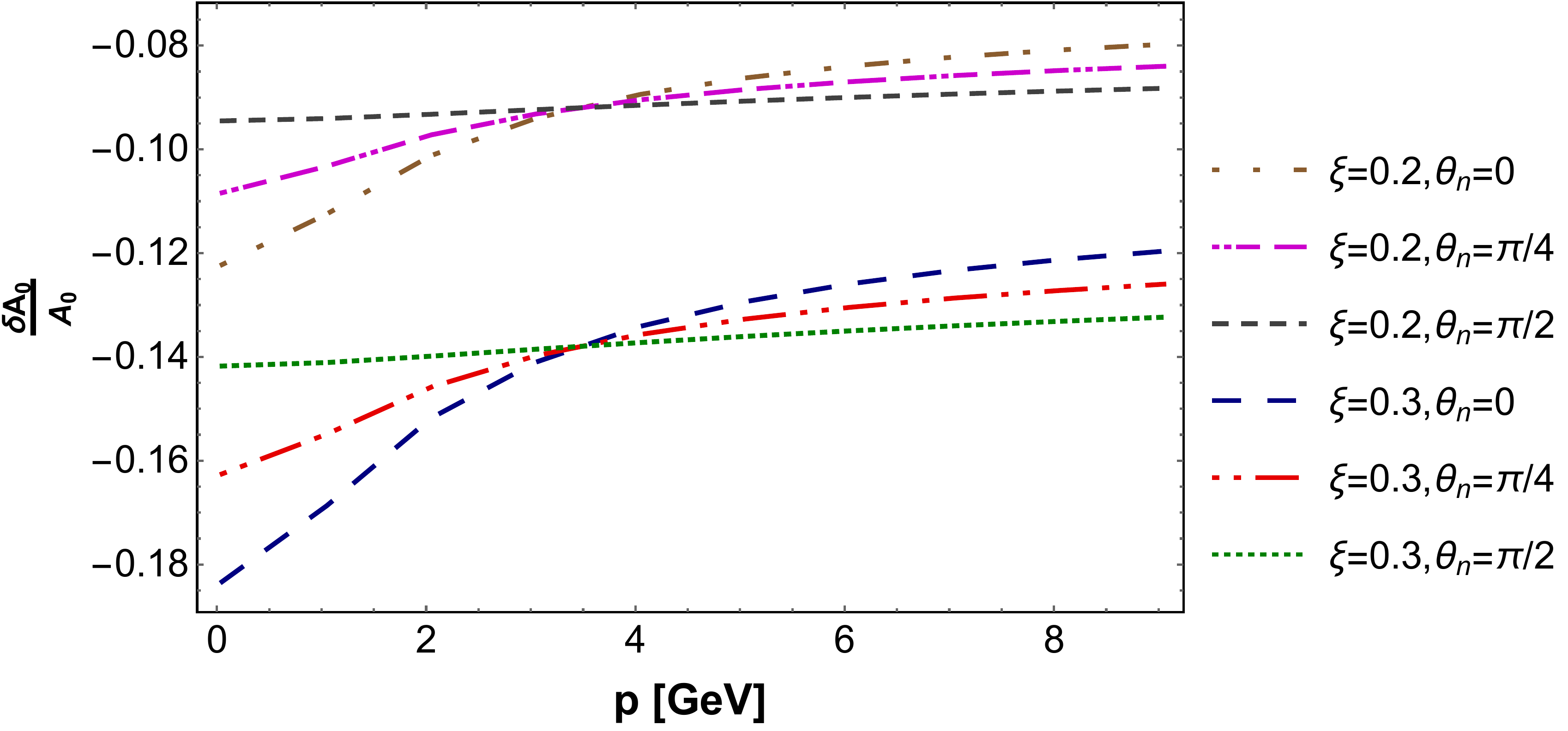}
    \includegraphics[width=0.48\textwidth]{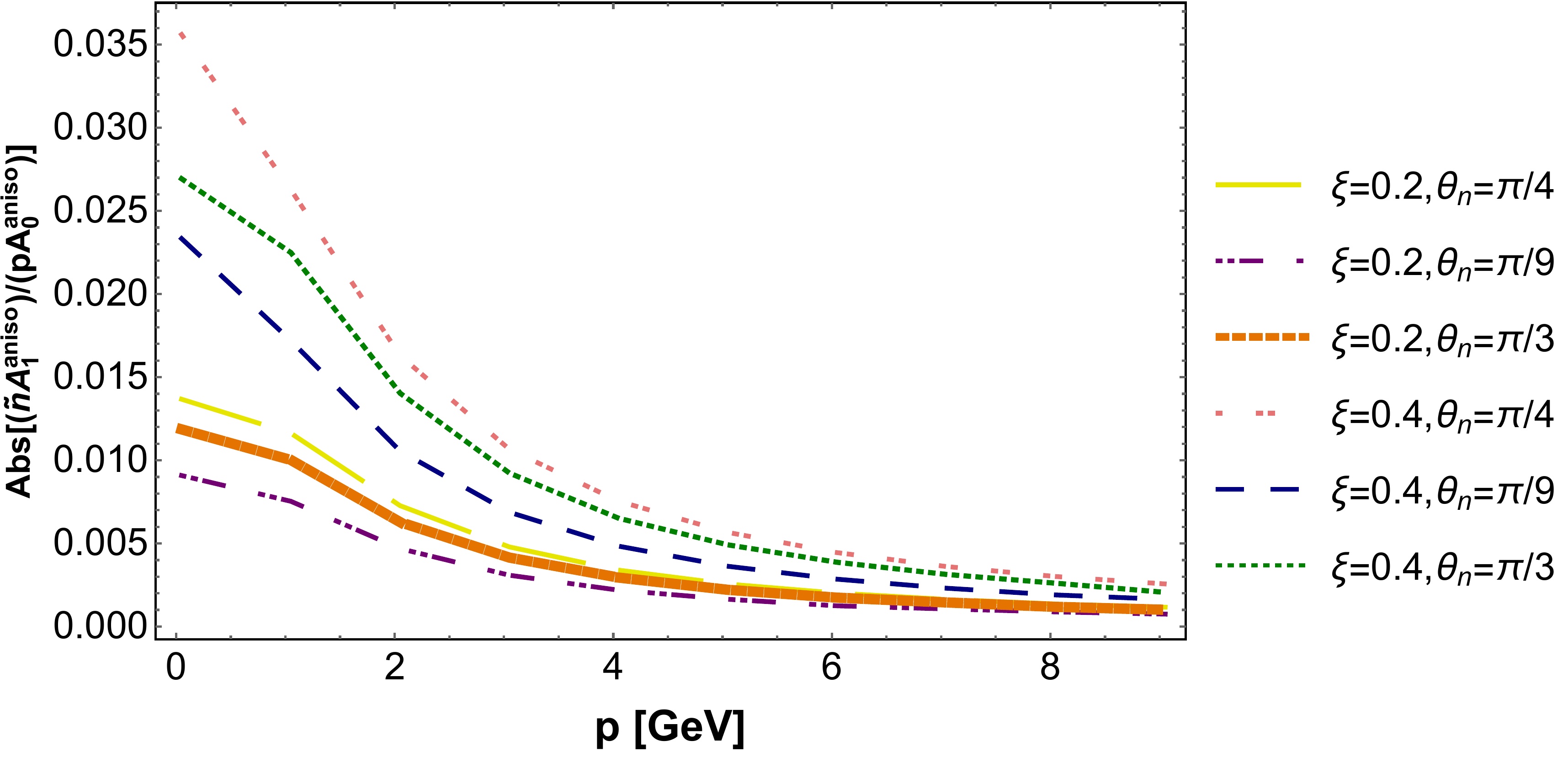}
    \includegraphics[width=0.48\textwidth]{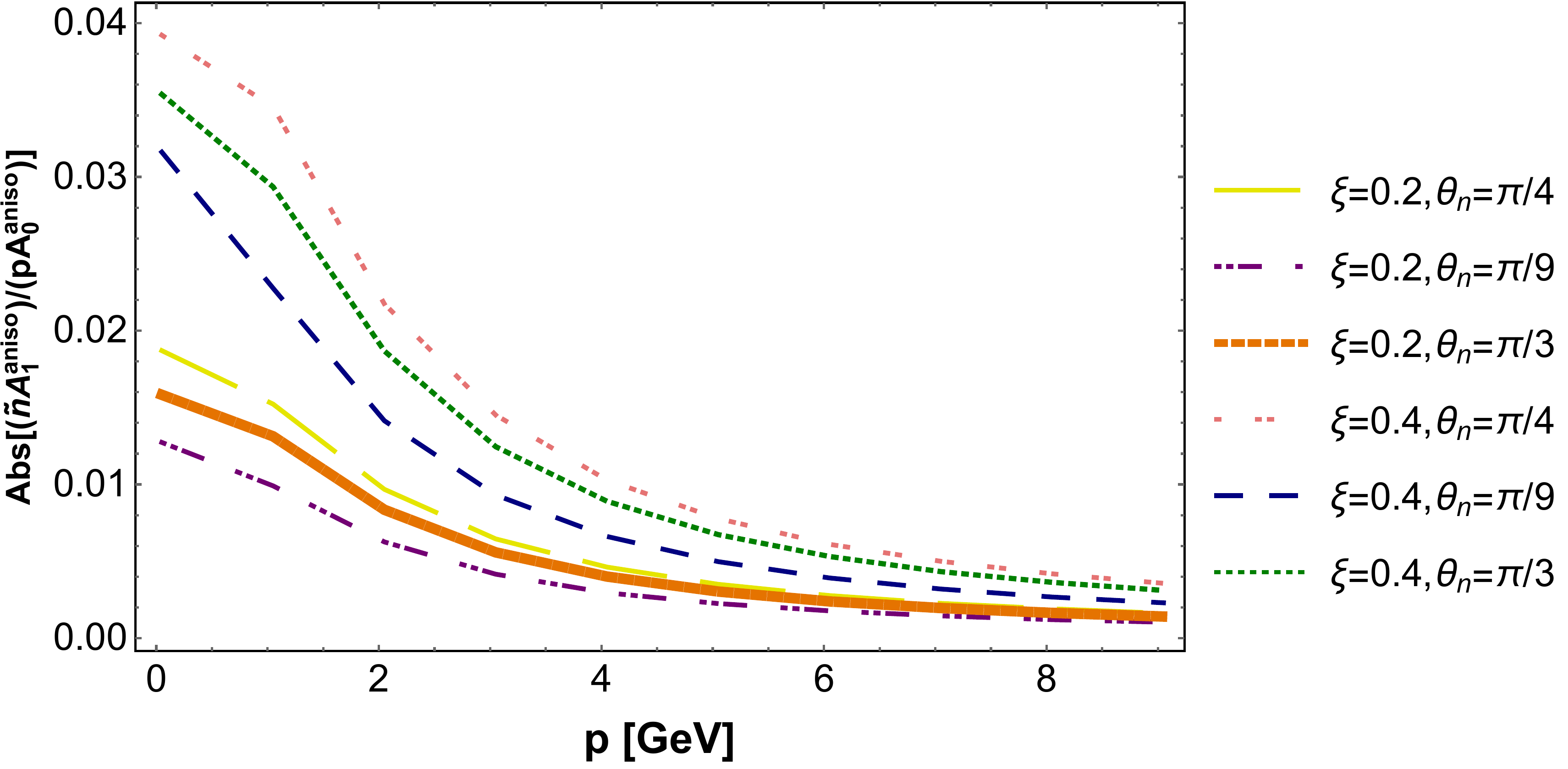}
    \caption{Anisotropic correction to $A_0$ as a function of its initial momentum at $T=360$ MeV (top panel). Relative significance of $A^{\text{(aniso)}}_1$ in comparison with $A^{\text{(aniso)}}_0$ at $T=360$ MeV (middle panel) and $T=480$ MeV (bottom panel).}
    \label{f1}
\end{figure}
Now, we proceed with the estimation of $\langle\langle {\bf{\Tilde{n}\cdot p'} \rangle\rangle}$ in center-of-mass frame. We have consider ${\bf {n}}=(\sin\theta_n, 0, \cos\theta_n)$, where angle $\theta_n$ is the angle between anisotropy vector and ${\bf \hat{z}}$. It is important to note that the analysis is also valid for the choice ${\bf {n}}=(0, \sin\theta_n, \cos\theta_n)$. Light quark momentum can be decomposed as ${\bf q} = (q\sin\chi\cos\phi, q\sin\chi\sin\phi, q\cos\chi)$ and HQ momentum chosen as ${\bf p} = (0, 0, p)$ such that we have,
\begin{align}\label{C.01}
&{\bf p}\cdot{\bf{q}} = pq\cos\chi,\\
& {\bf p}\cdot{\bf {n}} = p\cos\theta_n,\\
&{\bf q}\cdot{\bf {n}} =q\sin\chi\cos\phi \sin\theta_n+q\cos\chi\cos\theta_n,
\end{align}
We have $\Tilde{n}^i p^{'\,i}=p^{'\,i}\Big(\delta^{ij}-\dfrac{p^ip^j}{p^2}\Big)n^j$. Hence, we have 
\begin{align}
    \langle\langle {\bf{\Tilde{n}\cdot p'}}\rangle\rangle= \langle\langle{\bf{{n}\cdot p'}}\rangle\rangle- \langle\langle{\bf{{p}\cdot p'}}\rangle\rangle\frac{{\cos\theta_n}}{p}.
\end{align}
First, we need to obtain $({\bf{{n}\cdot p'}})$ in terms of other variables of integration. The Lorentz transformation that relates laboratory frame and center-of-mass frame has the following form,
\begin{align}
    &{\bf p}^{'}=\gamma_{cm}\Big({\bf \hat{ p}}_{cm}^{'}+{\bf v}_{cm}\hat{ E}_{cm}^{'}\Big),
\end{align}
where  {$\gamma_{cm}=\frac{E_p+E_q}{\sqrt{s}}$} and the velocity of the center-of-mass  {${\bf v}_{cm}=\frac{{\bf p}+{\bf q}}{E_p+E_q}$}. The energy conservation leads to $\hat{p}_{cm}^{'\, 2}=\hat{p}_{cm}^2$. In the center-of-mass frame, ${\bf  \hat{p}}_{cm}^{'}$ can be decomposed as follows,
\begin{align}\label{C.1}
   {\bf  \hat{p}}_{cm}^{'}=&\hat{p}_{cm}\Big(\cos\theta_{cm}{\bf \hat{x}}_{cm}+\sin\theta_{cm}\sin\phi_{cm}{\bf \hat{y}}_{cm}\nonumber\\ &+\sin\theta_{cm}\cos\phi_{cm}{\bf \hat{z}}_{cm}\Big),
\end{align}
where $\hat{p}_{cm}=\frac{s-m_{HQ}^2}{2\sqrt{s}}$ is the HQ momentum and $\hat{E}_{cm}=\sqrt{\hat{p}_{cm}^2+m^2_{HQ}}$ is the energy in the center-of-mass frame. The axis ${\bf  \hat{x}}_{cm}$, ${\bf  \hat{y}}_{cm}$, and ${\bf  \hat{z}}_{cm}$ are defined in Ref.~\cite{Svetitsky:1987gq}. Employing the above definitions, we obtain
\begin{widetext}
\begin{align}
   {\bf{\Tilde{n}\cdot p'}}=&\frac{\gamma_{cm}}{1+\gamma^2_{cm}v^2_{cm}}\Bigg\{{ \hat{ p}}_{cm}\bigg(\cos\theta_{cm}({\bf \hat{x}}_{cm}\cdot{\bf {n}})+ \sin\theta_{cm}\sin\phi_{cm}({\bf \hat{y}}_{cm}\cdot{\bf {n}} )+\sin\theta_{cm}\cos\phi_{cm}({\bf \hat{z}}_{cm}\cdot{\bf {n}} )\bigg)\nonumber\\
   &+\gamma_{cm}{E}^{'}_{p} \frac{(p\cos\theta_n+q\cos\chi\cos\theta_n+q\sin\chi\cos\phi \sin\theta_n)}{E_p+E_q}\Bigg\}-\frac{\gamma_{cm}}{1+\gamma^2_{cm}v^2_{cm}}\frac{\cos\theta_n}{p}\Bigg\{{ \hat{ p}}_{cm}\bigg(\cos\theta_{cm}({\bf \hat{x}}_{cm}\cdot{\bf {p}})\nonumber\\
   &+ \sin\theta_{cm}\sin\phi_{cm}({\bf \hat{y}}_{cm}\cdot{\bf {p}} )\bigg)+\gamma_{cm}{E}^{'}_{p}\frac{ (p^2+pq\cos\chi)}{E_p+E_q}\Bigg\}.
\end{align}
Note that we have obtained,
\begin{align*}
   {\bf{{p}\cdot p'}}&= \frac{\gamma_{cm}}{1+\gamma^2_{cm}v^2_{cm}}\Bigg\{{ \hat{ p}}_{cm}\bigg(\cos\theta_{cm}({\bf \hat{x}}_{cm}\cdot{\bf {p}})+ \sin\theta_{cm} \sin\phi_{cm}({\bf \hat{y}}_{cm}\cdot{\bf {p}} )\bigg)+\gamma_{cm}{E}^{'}_{p}\frac{ (p^2+pq\cos\chi)}{E_p+E_q}\Bigg\}\nonumber\\
  &=E_pE_p^{'}-\hat{E}_{cm}^2+\hat{p}_{cm}^2\cos\theta_{cm}.
  \end{align*}
\end{widetext}
The respective projections of the anisotropy vector and HQ momentum with the center-of-mass axis are defined in the Appendix~\ref{A}. 

\begin{figure}
    \centering
    \includegraphics[width=0.48\textwidth]{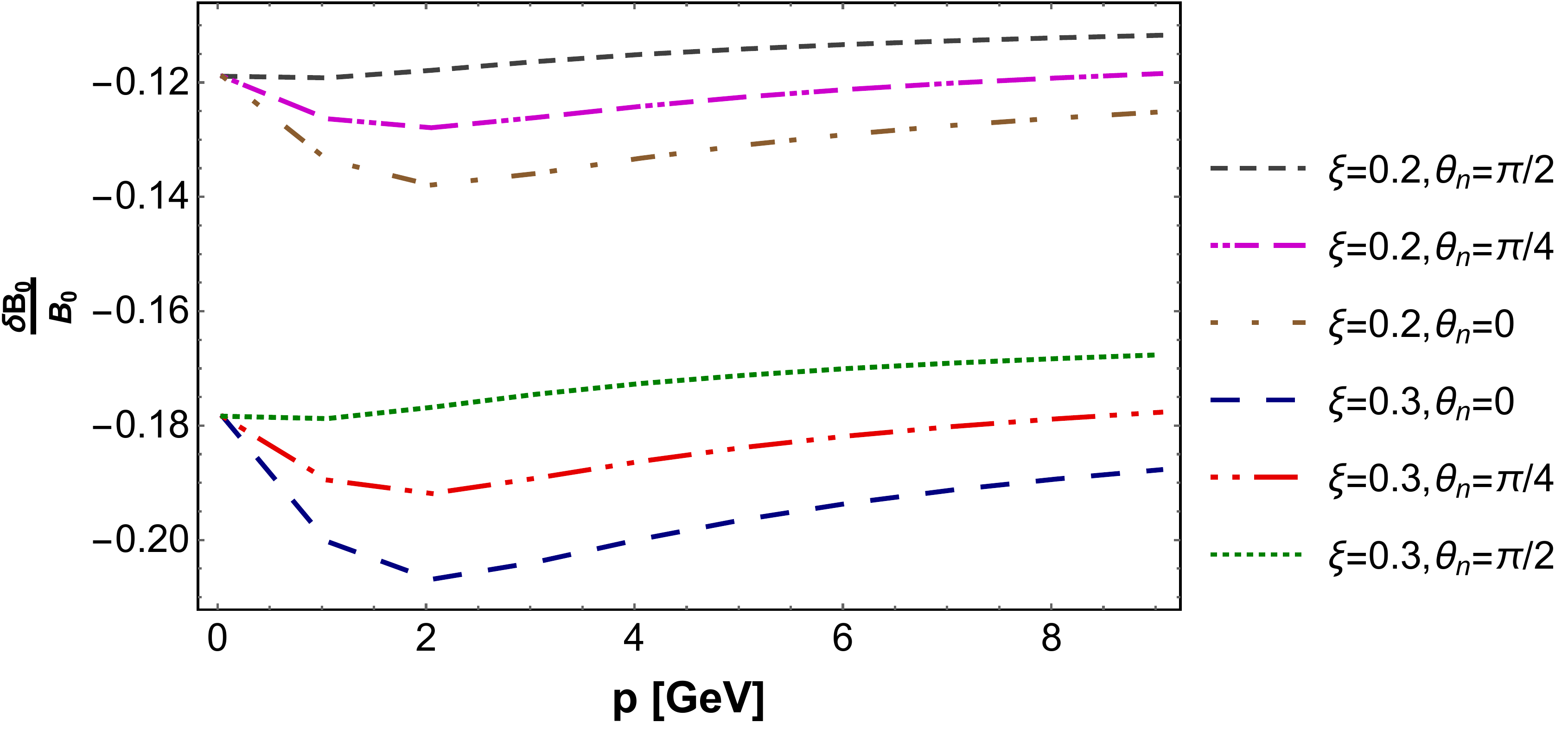}
    \includegraphics[width=0.48\textwidth]{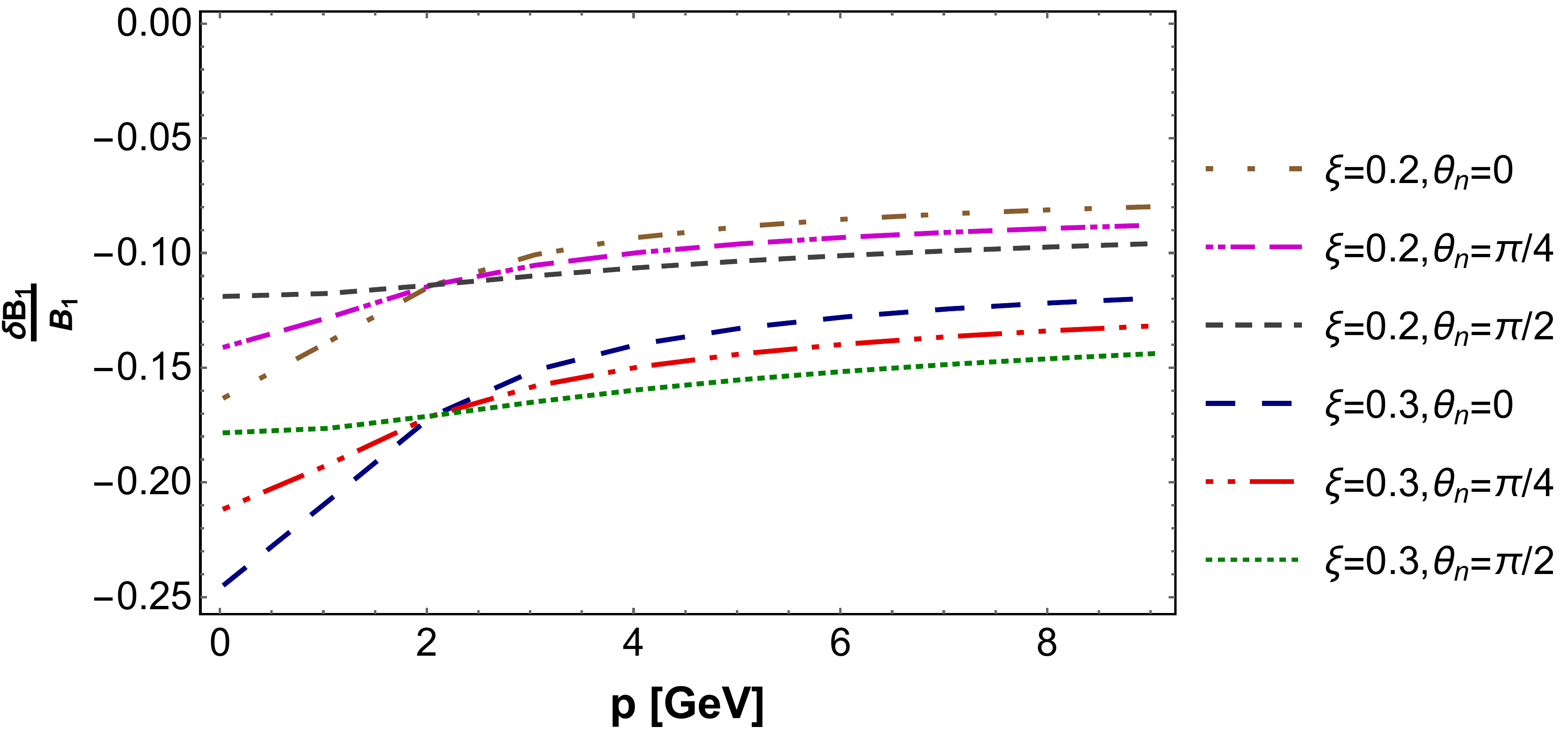}
    \caption{Momentum dependence of anisotropic corrections to $B_0$ (top panel) and $B_1$ (bottom panel) at $T=360$ MeV.}
    \label{f2}
\end{figure}
\section{Results and discussions}\label{sec:3}
\subsection{HQ transport coefficients in an anisotropic medium}
We initiate the discussions with the momentum dependence of the components of the HQ drag force in an anisotropic QCD medium. HQ drag force has two components, namely $A^{\text{(aniso)}}_0$ and $A^{\text{(aniso)}}_1$, in the anisotropic medium as described in Eq.~(\ref{1.13}).  The anisotropic effects are entering through the non-equilibrium part of the momentum distribution function. For the quantitative estimation, we have considered $m_{HQ}=1.3$ GeV for charm quarks, $a_k=0$, and one-loop running coupling constant from Ref.~\cite{Das:2015ana}. The impact of anisotropy on the momentum dependence of $A_0$ is depicted in Fig.\ref{f1} (top panel). The momentum and temperature of $A_0$ in the isotropic QCD medium have been well explored in Refs.~\cite{GolamMustafa:1997id,Cao:2018ews}.  The anisotropic part $\delta A_0$ considerably reduces the drag coefficient $A_0$, especially at the low momentum regimes. It is seen that the anisotropic correction has a strong dependence on the direction of anisotropy (with respect to the direction of HQ motion, $\theta_n$) and the strength of anisotropy in the medium. However, the dependence of the angle $\theta_n$ on the drag coefficient is observed to be opposite for low momentum regimes in comparison with the high momentum regimes. The additional drag coefficient $A^{\text{(aniso)}}_1$ arises due to the anisotropy of the medium. The relative significance of $A^{\text{(aniso)}}_1$ with that with $A^{\text{(aniso)}}_0$ is plotted as a function of HQ momentum at $T=360$ MeV and $T=480$ MeV in Fig.\ref{f1} (middle panel and bottom panel). We have observed that the additional component is negligible at the high momentum regimes. However, the additional drag coefficient may have a important role at the higher temperature regimes. It is important to note that the same decomposition of HQ drag force holds true in a strongly anisotropic medium, and the additional component may have a more visible impact on HQ motion with an increase in the strength of anisotropy as $\langle\langle {\bf{\Tilde{n}\cdot p'}}\rangle\rangle\propto \xi$.

The anisotropic corrections to the HQ diffusion coefficients $B^{\text{(aniso)}}_0=B_0+\delta B_0$ and $B^{\text{(aniso)}}_1=B_1+\delta B_1$ are plotted in Fig.~\ref{f2}. The impact of medium anisotropy is more pronounced in the low HQ momentum regimes. Unlike in the case of $B^{\text{(aniso)}}_1$, the coefficient $B^{\text{(aniso)}}_0$ gets anisotropic contribution from $\langle\langle ({\bf{p'\cdot \Tilde{n}}})^2\rangle\rangle$ along with non-equilibrium part of the thermal distribution function. In the limit $\xi\rightarrow 0$, the forms of $B^{\text{(aniso)}}_0$ and $B^{\text{(aniso)}}_1$ as described in Eq.~(\ref{B0.0}) and Eq.~(\ref{B0.1})  will reduce back to the results of Ref.~\cite{Svetitsky:1987gq} (if we use same parameters as used in Ref.~\cite{Svetitsky:1987gq}). Both the momentum behaviour of diffusion coefficients are seen to have a strong dependence on the angle between the anisotropic vector and HQ velocity in the QCD medium.

\begin{figure}
    \centering
    \hspace{.5cm}
    \includegraphics[width=0.45\textwidth]{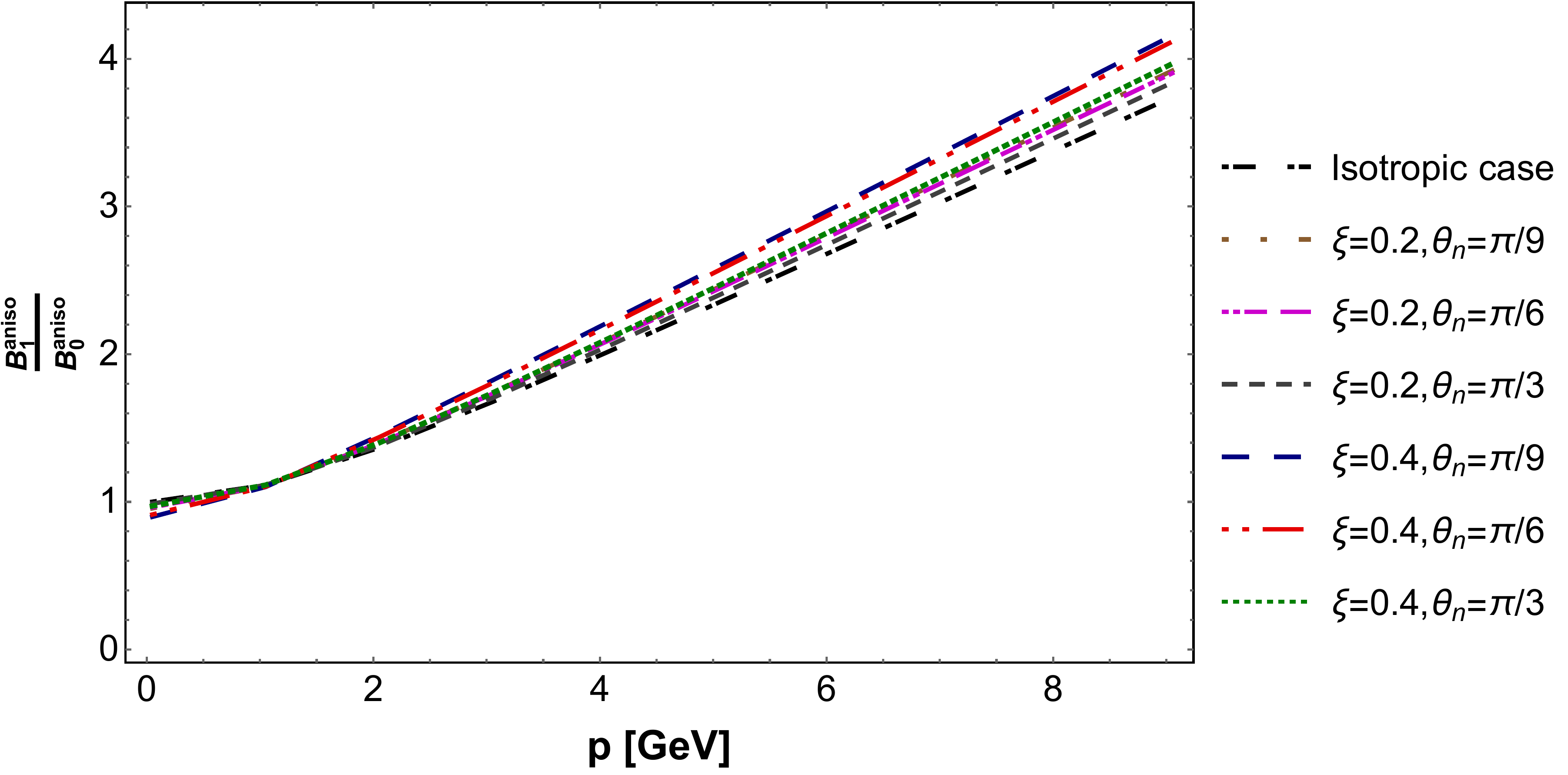}
    \includegraphics[width=0.50\textwidth]{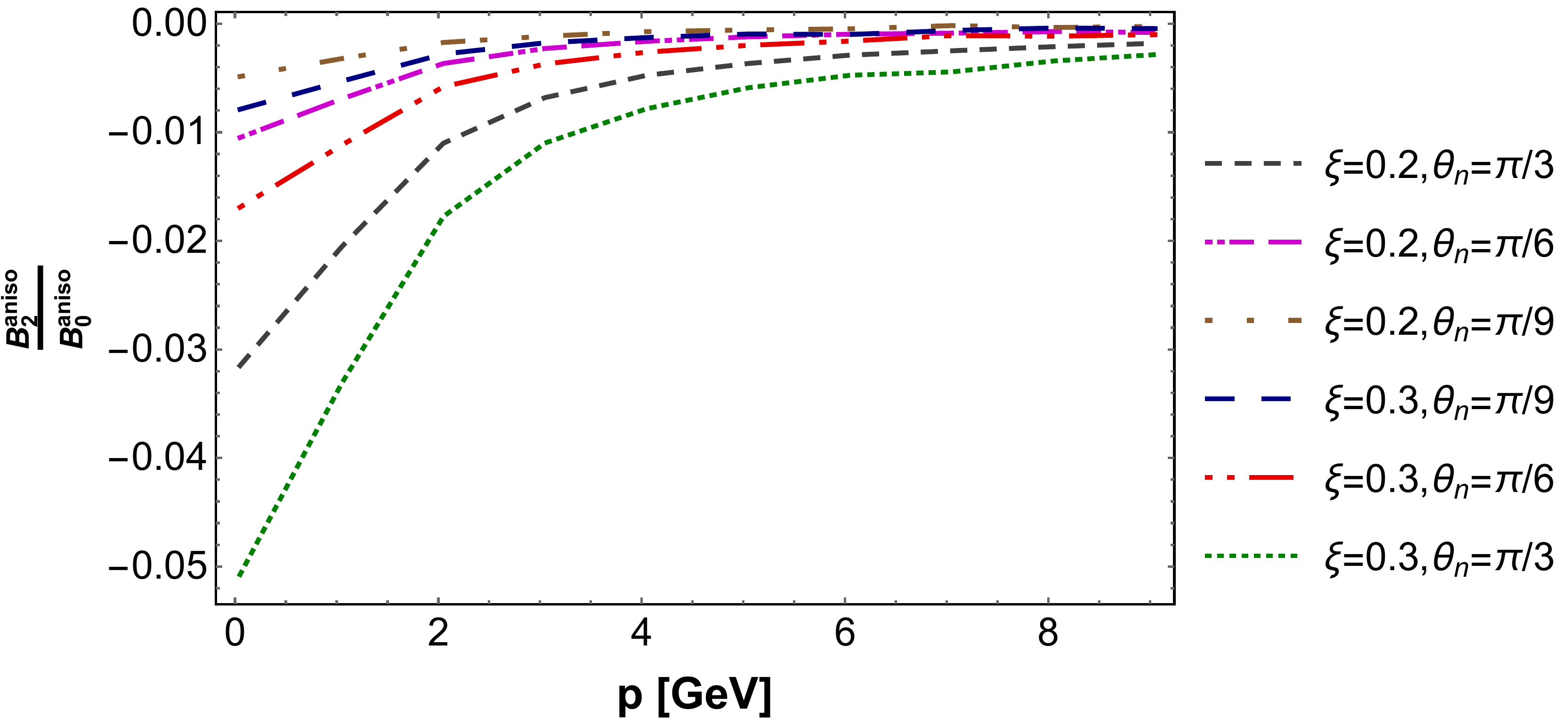}
    \includegraphics[width=0.50\textwidth]{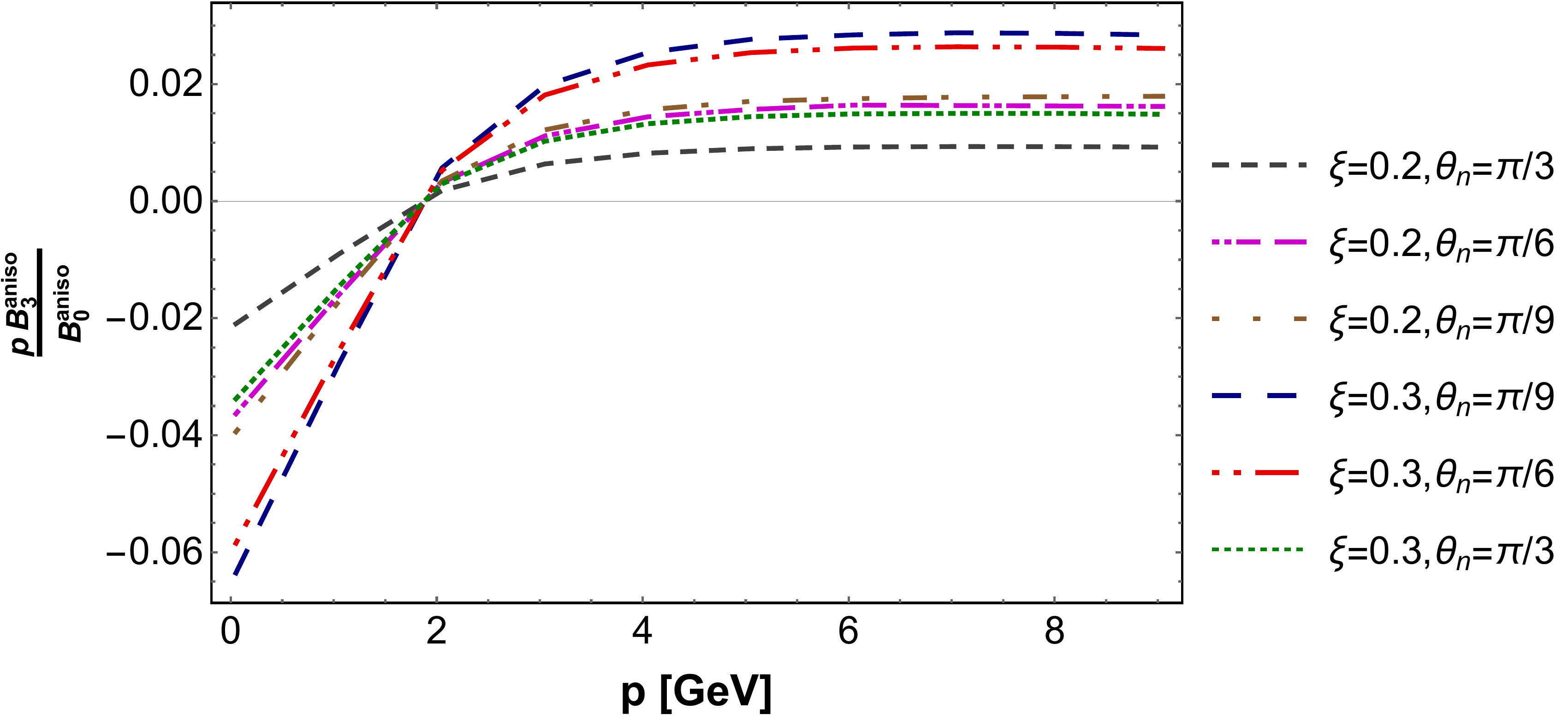}
    \caption{ Relative significance of HQ diffusion coefficients in an anisotropic medium:  ${B^{\text{(aniso)}}_1}/{B^{\text{(aniso)}}_0}$ (top panel), ${B^{\text{(aniso)}}_2}/{B^{\text{(aniso)}}_0}$ (middle panel), $(p{B^{\text{(aniso)}}_3})/{B^{\text{(aniso)}}_0}$ (bottom panel) at $T=360$ MeV.}
    \label{f3}
\end{figure}
The momentum anisotropy in the medium further give rise to additional components of HQ diffusion coefficients, namely $B^{\text{(aniso)}}_2$ and $B^{\text{(aniso)}}_3$. Note that these additional coefficients vanish in the isotropic limit as $\xi\rightarrow 0$. The relative significance of HQ diffusion coefficients in an anisotropic QCD medium is depicted in Fig.~\ref{f3}. The additional diffusion coefficients seem to be more prominent in the low momentum regimes in comparison with high momentum regimes. In the static limit $p\rightarrow 0$, these coefficients are non-negligible, especially for the case of a strongly anisotropic medium.  However, the coefficient $B^{\text{(aniso)}}_1$ is dominant over $B^{\text{(aniso)}}_0$ at high momenta. Similar to the case of isotropic medium, in the static limit, we obtain $B^{\text{(aniso)}}_0=B^{\text{(aniso)}}_1$.
The direction of anisotropy in the medium has a visible impact on the low momentum behaviour of $B^{\text{(aniso)}}_2$ and $B^{\text{(aniso)}}_3$. Whereas the angle $\theta_n$ dependence is negligible for the momentum behaviour of $B^{\text{(aniso)}}_1$.
\subsection{HQ energy loss in an anisotropic medium}

\begin{figure}
    \centering
    \includegraphics[width=0.48\textwidth]{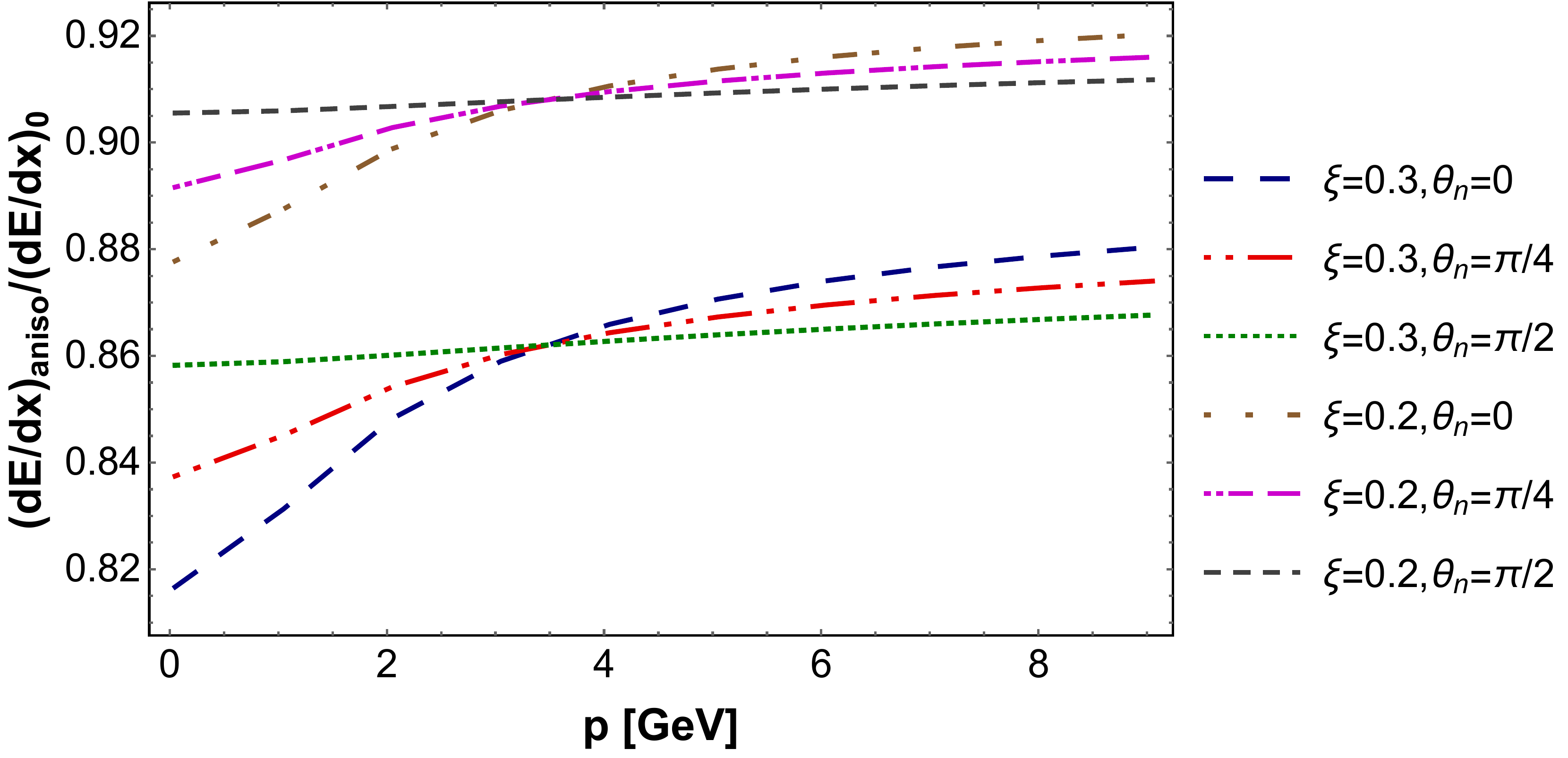}
    \includegraphics[width=0.48\textwidth]{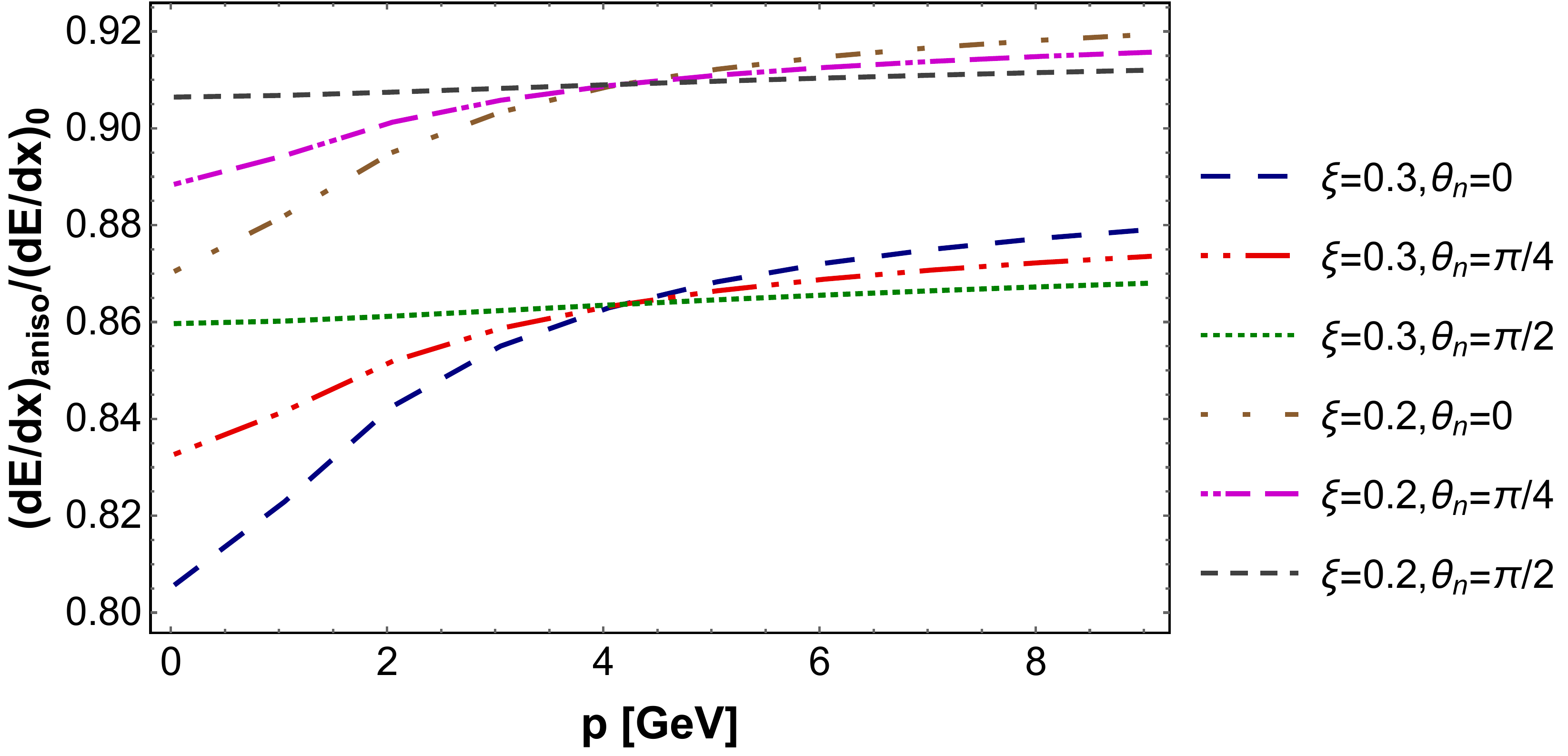}
    \caption{Impact of anisotropy on the momentum behaviour of collisional energy loss of charm quark for the RHIC energy at $T=360$ MeV (top panel) and for the LHC energy at $T=480$ MeV.}
    \label{f4}
\end{figure}

HQs may lose its energy while traveling through the anisotropic QCD medium due to the collisional processes with the in-medium particles. The differential collisional energy loss can be quantified in terms of the HQ drag coefficient due to the elastic collisions in the medium as~\cite{GolamMustafa:1997id},
\begin{equation}\label{37}
\Big(-\dfrac{dE}{d\text{x}}\Big)_{\text{aniso}}=A^{\text{(aniso)}}_0(p^2, T)p.
\end{equation}
It is important to note that the current focus is on the energy loss in the direction of initial HQ momentum. Hence, the contribution from $A^{\text{(aniso)}}_1$ will vanish as ${\bf p}\cdot{\bf \Tilde{n}}=0$. However, the energy loss will have an anisotropic contribution through the $\delta A_0$. We have plotted the ratio of collisional energy loss of charm quark in an anisotropic medium $(-\frac{dE}{d\text{x}})_{\text{aniso}}$ to that in the isotropic QCD medium $(-\frac{dE}{d\text{x}})_0$ for the RHIC and LHC energies in Fig.~\ref{f4}. The quark energy loss of HQ in the QCD medium seems to have a dependence on its initial momentum and temperature of the background medium. The energy loss of HQ gets suppressed in the anisotropic QCD medium with an increase in the strength of the anisotropy factor. However, the direction of anisotropy in the medium seems to have a weaker dependence on the HQ energy loss for the RHIC energy. These observations on charm quark energy loss hold true for the LHC energy too.  

\section{Conclusion and outlook}\label{sec:4}
We have studied the HQ transport coefficients and energy loss in an anisotropic QCD medium within the Fokker-Planck approach. The anisotropic aspect of the medium has been incorporated in the analysis through the non-equilibrium part of the quarks/antiquarks and gluonic momentum distribution. We have employed a proper decomposition to HQ drag force with two components in the anisotropic QCD medium. Similarly, we have constructed the second rank HQ diffusion tensor with four diffusion coefficients in the anisotropic medium.  We have realized that the anisotropic effects have a strong dependence on the orientation of HQ motion with the direction of anisotropy in the medium. The relative significance of these anisotropic transport coefficients has been studied as a function of HQ initial momentum. It is seen that the additional components of drag and diffusion coefficients that arise due to the momentum anisotropy of the medium are sub-dominant in comparison with the isotropic components for weakly anisotropic medium. Moreover, these anisotropic contributions are essential for the theoretical consistency for studying the HQ transport in an anisotropic QCD medium. Further, we have analyzed the anisotropic contribution to HQ collisional energy loss. It is observed that the HQ energy loss depends on the relative orientation of anisotropy with the HQ motion, especially in the low momentum regimes. 

The anisotropic drag and momentum diffusion can be used as input parameters for the study of experimentally measured observables associated with HQs in heavy-ion collisions, such as flow coefficients and nuclear suppression factor. The inclusion of the impact of momentum anisotropy of the medium to the HQ may have a visible effect on the experimental observables. We intend to study the phenomenological aspect of the anisotropic HQ transport coefficients in follow-up work. It is an interesting task to study HQ dynamics in a strongly anisotropic medium. Perhaps, the additional drag and diffusion coefficients due to the anisotropy in the medium may have a significant role in a strongly anisotropic medium. We intend to explore this aspect in the near future.  The recent LHC and  RHIC findings on the enhanced directed flow of D-meson~\cite{ALICE:2019sgg,STAR:2019clv} give the indications on the existence of a magnetic field generated in heavy-ion collisions. The magnetic field induces anisotropy in the system and gives rise to field-induced anisotropic HQ drag and momentum diffusion coefficients in a magnetized medium. The radiative process by HQs (inelastic process) in the anisotropic magnetized QCD medium is another interesting direction to explore. 
\section*{Acknowledgments}
V. C. and S. K. D. acknowledge the SERB Core Research Grant (CRG) [CRG/2020/002320].  M. K. and A. K.  acknowledge the Indian Institute of Technology Gandhinagar for the Institute postdoctoral fellowship. 

\appendix
\section{ Projections of anisotropy vector and HQ momentum with the center-of mass axis } \label{A}
By employing the following definitions in the center-of-mass frame,

 {
\begin{align}
  ({\bf v}_{cm}\cdot {\bf  \hat{p}}_{cm})&=\gamma_{cm}\Bigg[\frac{(p^2+pq\cos\chi)}{E_p+E_q}-v^2_{cm}{E}_{p}\Bigg],\\
  N^2&=v^2_{cm}-\frac{({\bf v}_{cm}\cdot {\bf  \hat{p}}_{cm})^2}{\hat{p}^2_{cm}},\\
  v^2_{cm}&=\frac{p^2+q^2+2pq\cos\chi}{(E_p+E_q)^2},
\end{align}
} 
we have, 
\begin{widetext}
 {
\begin{align}
   ({\bf \hat{x}}_{cm}\cdot{\bf {n}})&=\frac{\gamma_{cm}}{\hat{p}_{cm}}\Bigg[p\cos\theta_{n}-{E}_p\frac{(p\cos\theta_{n}+q\cos\chi\cos\theta_n+q\sin\chi\cos\phi \sin\theta_n)}{E_p+E_q}\Bigg],\\
   ({\bf \hat{y}}_{cm}\cdot{\bf {n}})
   &=N^{-1}\Bigg[\frac{(p\cos\theta_{n}+q\cos\chi\cos\theta_n+q\sin\chi\cos\phi \sin\theta_n)}{E_p+E_q}\nonumber\\
   &~~~~~-({\bf v}_{cm}\cdot {\bf  \hat{p}}_{cm})\frac{\gamma_{cm}}{\hat{p}^2_{cm}}\bigg(p\cos\theta_{n} -{E}_{p}\frac{(p\cos\theta_{n}+q\cos\chi\cos\theta_n+q\sin\chi\cos\phi \sin\theta_n)}{E_p+E_q}\bigg)\Bigg],
   \end{align}
   \begin{align}
   ({\bf \hat{z}}_{cm}\cdot{\bf {n}})&={\gamma_{cm}}N^{-1}\frac{1}{\hat{p}_{cm}(E_p+E_q)}p q \sin\chi \sin\phi\sin\theta_n,\\
   ({\bf \hat{x}}_{cm}\cdot{\bf {p}})&=\frac{\gamma_{cm}}{\hat{p}_{cm}}\Bigg[p^2-{E}_{p}\frac{(p^2+pq\cos\chi)}{E_p+E_q}\Bigg],\\
   ({\bf \hat{y}}_{cm}\cdot{\bf {p}})&=N^{-1}\Bigg[\frac{(p^2+pq\cos\chi)}{E_p+E_q}-({\bf v}_{cm}\cdot {\bf  \hat{p}}_{cm})\frac{\gamma_{cm}}{\hat{p}^2_{cm}}\bigg(p^2 -{E}_{p}\frac{(p^2+pq\cos\chi)}{E_p+E_q}\bigg)\Bigg].
\end{align}
}
\end{widetext}

\bibliography{pv_ref}{}
\end{document}